\documentclass[12pt]{article}

\usepackage{scicite}

\usepackage{amsmath,comment}
\usepackage{times}
\usepackage{todonotes}

\newenvironment{sciabstract}{%
\begin{quote} \bf}
{\end{quote}}

\newcounter{lastnote}

\title{Relatedness in the Era of Machine Learning}

\author
{Andrea Tacchella$^{1}$, Andrea Zaccaria$^{2,3\ast}$, Marco Miccheli$^{2,4,5}$, Luciano Pietronero$^{3}$\\
\\
\normalsize{$^{1}$Joint Research Centre, Seville}\\
\normalsize{$^{2}$Istituto dei Sistemi Complessi - CNR, UOS Sapienza, Rome}\\
\normalsize{$^{3}$Centro Ricerche Enrico Fermi, Rome}\\
\normalsize{$^{4}$Translated Srl, Rome}\\
\normalsize{$^{5}$Dipartimento di Fisica, Università Sapienza, Rome}\\
\normalsize{$^\ast$To whom correspondence should be addressed; E-mail:  andrea.zaccaria@cnr.it}
}

\date{}

\begin{document} 


\baselineskip24pt


\maketitle


\begin{sciabstract}
  Relatedness is a quantification of how much two human activities are similar in terms of the inputs and contexts needed for their development. Under the idea that it is easier to move between related activities than towards unrelated ones, empirical approaches to quantify relatedness are currently used as predictive tools to inform policies and development strategies in governments, international organizations, and firms. Here we focus on countries' industries and we show that the standard, widespread approach of estimating Relatedness through the co-location of activities (e.g. Product Space) generates a measure of relatedness that performs worse than trivial auto-correlation prediction strategies. We argue that this is a consequence of the poor signal-to-noise ratio present in international trade data. In this paper we show two main findings. First, we find that a shift from two-products correlations (network-density based) to many-products correlations (decision trees) can dramatically improve the quality of forecasts with a corresponding reduction of the risk of wrong policy choices. Then, we propose a new methodology to empirically estimate Relatedness that we call Continuous Projection Space (CPS). CPS, which can be seen as a general network embedding technique, vastly outperforms all the co-location, network-based approaches, while retaining a similar interpretability in terms of pairwise distances. 
\end{sciabstract}


\section*{Introduction}
The concept of Relatedness \cite{hidalgo2018principle} is a key element for both economic and social sciences, with applications ranging from smart specialization strategies \cite{balland2019smart} to the study of countries development \cite{hidalgo2007product,zaccaria2014taxonomy}, to recommender systems \cite{lu2012recommender}.
Two human activities are considered to be \textit{related} if they share a common set of capabilities that are needed for their development\cite{teece1994understanding}. The larger is the intersection of the needed capabilities, the stronger is the Relatedness. For example, industrial production of radars is closely related to the production of radio broadcasting apparatus, but not to so much to crude oil refining.
In recent years, driven by the increasing popularity and adoption of the Economic Complexity framework \cite{hidalgo2021economic,hidalgo2009building,hidalgo2007product,Tacchella,tacchella2013economic,cristelli2013measuring,zaccaria2014taxonomy,sbardella2018role}, Relatedness has been gaining importance in informing diversification or specialization strategies across a wide range of policy making institutions such as the World Bank \cite{lin2020african} and the European Commission \cite{pugliese2019economic,pugliese2019economic2}. Under the idea that it is easier to develop activities that are similar to those already developed in a region, decision makers can rely on a quantitative tool to design policies that can be adapted to several strategic approaches (e.g. vertical or horizontal policies \cite{van2020vertical}). So a country, or a region, that is currently competitive in the production of radars can have development opportunities in radio broadcasting apparatus, as a consistent set of the needed capabilities is likely already present. With this paper we address the problem of providing a reliable quantification of the feasibility of this transition, that can be a valuable input for strategic decisions.

Clearly, a poor or inconsistent quantification of Relatedness, and therefore a wrong estimation of the feasibility of transitions to new industries, represents a huge risk for a policy maker basing its decisions on it. In fact, despite the potentially great impact that such ideas can have in shaping policies, an important point that needs to be addressed is the fact that there is no direct way to estimate Relatedness of real-world activities from first principles, i.e. through the quantification of common inputs. While some theoretical work has been done on the combinatorics of very specific or synthetic networks where the inputs layer is observed (such as letters-words or ingredient-recipes networks \cite{zaccaria2014taxonomy,tacchella2016build}), in any real scenario involving human activities (e.g. industries, technologies, scientific research, etc.) we do not have access to any 'book of recipes', not even to the 'list of possible ingredients', that would allow for a principled computation of Relatedness in terms of shared input capabilities.

For this reason research has been focusing on how to recover an effective measure of Relatedness from location-activities data \cite{teece1994understanding,hidalgo2007product,neffke2011regions, zaccaria2014taxonomy}. The core idea is that if two activities require similar inputs they tend to co-occur within the same locations more than randomly\cite{teece1994understanding,saracco2017inferring,pugliese2019unfolding}. Therefore, suitably normalized counts of co-occurrences can be used as a proxy for Relatedness.
The problem that we address with this paper is that this proxied estimation is inherently difficult for two reasons:
\begin{itemize}
    \item The number of activities is very often much larger than the number of locations in which to count co-occurrences. This means that the correlation structure that emerges is mostly random. E.g. in the countries-products case, one would estimate a 5000*5000 (Harmonized System 6-digits) co-occurrence matrix of products out of approximately 170 observations (countries). It is possible to reduce the number of activities by aggregation \cite{zaccaria2018integrating,stojkoski2016impact}, but this typically leads to a very coarse grained Relatedness structure that is often trivial and of no practical interest. Also the opposite approach, namely to increase the number of locations by reducing granularity (i.e. going at the subnational level), is of little use due to the fact that harmonised regional-level data is often unavailable, and the number of observations needed to produce a good estimate of a 5000*5000 correlation matrix is easily in the tens or hundreds of thousands\cite{bun2017cleaning}.
    \item Very often, locations-activities bipartite networks have a very strong nested structure \cite{mariani2019nestedness}. As opposed to a block-diagonal structure, that would immediately lead to a definition of sector-communities, in these networks the Relatedness signal is of second order with respect to the drive towards diversification that generates the nested structure. 
\end{itemize}


The basic idea of counting co-occurrences to infer Relatedness has been refined and generalized in a wide variety of approaches \cite{teece1994understanding,zaccaria2014taxonomy,nesta2005coherence,bottazzi2010measuring,li2014statistically,saracco2017inferring,cimini2019statistical,giorgio2020knowledge,zhou2010solving,zhou2007bipartite,pugliese2019unfolding}. All such approaches give rise to a network of Relatedness relations between couples of activities, i.e., in the language of statistical physics, a two-bodies correlation structure. 

\begin{figure}[h!]
\centering
\includegraphics[scale=0.75]{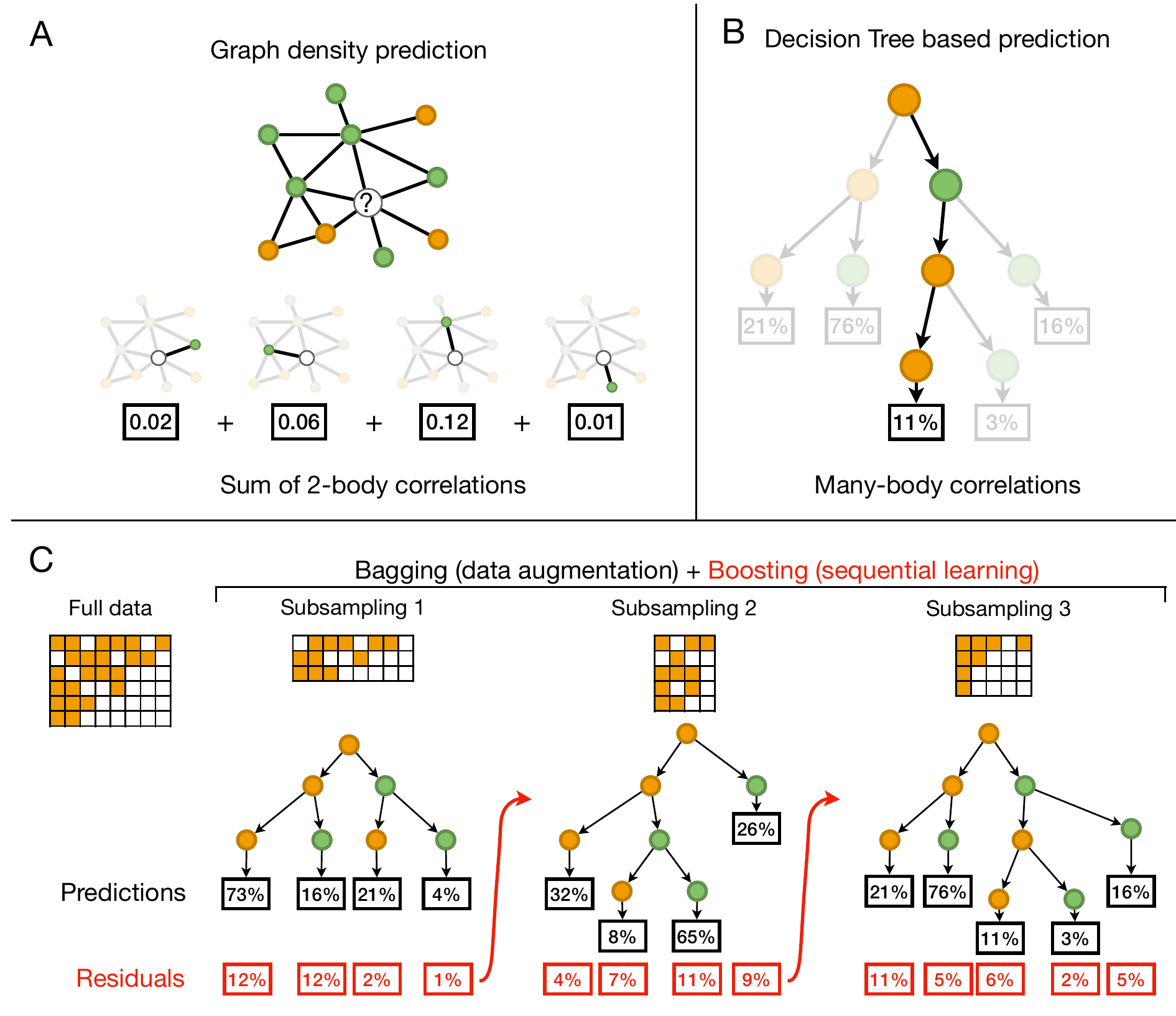}
\caption{\textbf{A} Density-based predictions on a graph are a linear sum of two-products relations. \textbf{B.} Decision-Tree based predictions are based on many-products relations, i.e. the full path on the decision tree. A change in the value of one of the nodes has a non-linear effect on the prediction. \textbf{C.} Visual explanation of the Bagging and Boosting paradigms implemented in the XGBoost tree-based models. Bagging: The full training data is sub-sampled (for the sake of visualization here we represent one subsampling per tree, but can be done at the node level) and different \textit{weak} models are learned on each subsample. Boosting: each model is trained to optimize the residuals of the averaged previous models.}
\label{fig:trees_vs_graphs}
\end{figure}

In order to give an objective assessment of the quality of these proxies for Relatedness, we test the ability of these networks to perform an out-of-sample link prediction task in the countries-products bipartite network. We perform the tests in a cross-validation setting that is detailed in the Methods section. The finding is that the link prediction ability of the co-occurrence methods is generally poor, in some cases only marginally better than the one produced by a random network, and always inferior to trivial prediction strategies. From this evidence one could draw two kinds of conclusions: either (i) Relatedness is an unimportant concept in predicting the development paths of countries, or (ii) the co-occurrence proxy is able to provide a poor quantification of Relatedness.
Our findings are strongly in favour of (ii). While we find very poor link-prediction performances from co-occurrence based topologies, at the same time we are able to build much better Relatedness proxies through more advanced machine learning based embedding techniques (see Methods). However, we also find that describing countries development paths as the sum of binary Relatedness relationships is an oversimplification, and that much better results can be obtained considering higher-order interactions (i.e. patterns of absence/presence of many activities) through more complex but less interpretable tree-based models (fig. \ref{fig:trees_vs_graphs} panels A and B). These models can be learned with very effective strategies of \textit{data augmentation} (bagging), where many models are learned on randomized subsamples of data and then averaged, and \textit{boosting}, where the models are learnt in sequence, with each new model trained to minimise the residuals of the previous ones (fig. \ref{fig:trees_vs_graphs} panel C). These strategies allow to better cope with the relative scarcity of data and to learn more complex and effective models. Finally, we mention early attempts to adopt machine learning approaches in economic complexity \cite{che2020intelligent,tacchella2020language,palmucci2020your}, which however either lack the systematic comparison in prediction tasks we show here, use different data, discuss only very specific test cases, or propose methodologies which are not suitable for the type and amount of data relevant here.\\
In summary, in this paper we show that the current approaches for the estimation of Relatedness are often too simplistic and unable to provide reliable quantifications, and in particular that they are often only marginally better than a random signal for prediction tasks and always worse than trivial strategies. We perform an extensive testing and comparison of several methods and we propose a novel approach, that we call Continuous Projection Space, that dramatically improves prediction performances (+259\%  precision of top 1000 recommendations, +46\% BestF1 Score with respect to the Product Space baseline) while retaining a straightforward two-products Relatedness interpretability. Moreover, by shifting to a more complex, many-products Relatedness model with machine learning techniques, we are able to further substantially improve the quality of the predictions (+350\%  precision of top 1000 recommendations, +96\% BestF1 Score with respect to the Product Space baseline).

\section*{Results}
\paragraph*{Definition of the problem}
 We consider the temporal countries-product network defined by the bipartite adjacency matrix
$$M_{cp}(t) = 
\begin{cases}
1, & \text{if } RCA_{cp}(t) \geq 1\\
0, & \text{otherwise}
\end{cases}$$
where $RCA_{cp}(t)$ is the Revealed Comparative Advantage\cite{balassa1965trade} of country $c$ in product $p$ in year $t$, based on COMTRADE data (see Methods section). 
Roughly speaking, $M_{cp}=1$ means that country $c$ is exporting $p$ in a competitive way.\\
We want to test if co-occurrence based methods are a good proxy for the Relatedness of countries' traded goods. In order to provide a quantitative and objective evaluation of how good these proxies are, we make use of the standard assumption that countries are more likely to develop new products that are related to the ones they already produce \cite{hidalgo2007product,zaccaria2014taxonomy,alshamsi2018optimal}. Therefore, our validation criteria are all related to the ability of any inferred Relatedness topology to predict the export basket of a country in year $t+\delta$ given its export basket in year $t$. Here we discuss the case $\delta=5$, however different values of $\delta$ give similar results. More precisely, we implement a \emph{leave-k-countries-out} cross-validation strategy (see Methods section), so that we learn both Relatedness topologies and predictive models from a set of countries and then we test them on a different, non-overlapping set of countries. 
We consider various classes of predictive models, whose exact specifications are given in the Methods section:
\begin{itemize}
   \item Baselines: no co-occurrence models that completely disregard or randomize the co-occurence signal. The RCA method is based on the auto-correlation of the data.
    \item Bipartite Projections: inference of a Relatedness graph based on the bipartite countries-products structure. Co-occurrence based techniques such as the Product Space \cite{hidalgo2007product} and the Taxonomy Network \cite{zaccaria2014taxonomy} belong to this class.
    \item Description-based: Relatedness topology here is based on the textual similarity \cite{fuzzyset} between product descriptions in the Harmonized System
    \item Tree Based: tree-based machine learning algorithms that make a link prediction based on complex patterns of presence-absence of many links (many-body correlations). In particular, here we use XGBoost \cite{friedman2001greedy,chen2016xgboost} (XGB).
    \item CPS: low-dimensional representations of suitable embeddings obtained from the supervised machine learning algorithms above. We used TSNE \cite{van2008visualizing} and Variational Auto-Encoders \cite{kingma2013auto} for the dimensionality reduction.
\end{itemize}

We test the predictive models on two link prediction tasks, namely (1) to predict all the country-products links at time $t+\delta$ and (2) to predict the links at time $t+\delta$ that had $RCA(t) < 0.25$. The first task is much less interesting from an economic point of view because of the very strong auto-correlation of the countries-products structure: a model that trivially predicts that $M_{cp}(t+\delta) = M_{cp}(t)$ is typically able to achieve very high scores with every classification metric. On the other hand, the second task measures the ability of the predictive models to forecast new links that are not due to small fluctuations of RCA, i.e. that are more likely to represent genuine economic development. Here we therefore focus on the results of the experiments on task 2. For the sake of completeness, we report in the SI also the results for task 1. We perform the experiments using 12 years of data, from 2007 to 2018. In particular, for task 2 we compute a prediction score for each 0 element of the $M_{cp}$ matrix from 2007 to 2013 and we validate the predictions by checking whether those links are present 5 years later, i.e. in the 2012-2018 data.


\begin{figure}[htb]
\centering
\includegraphics[scale=0.54]{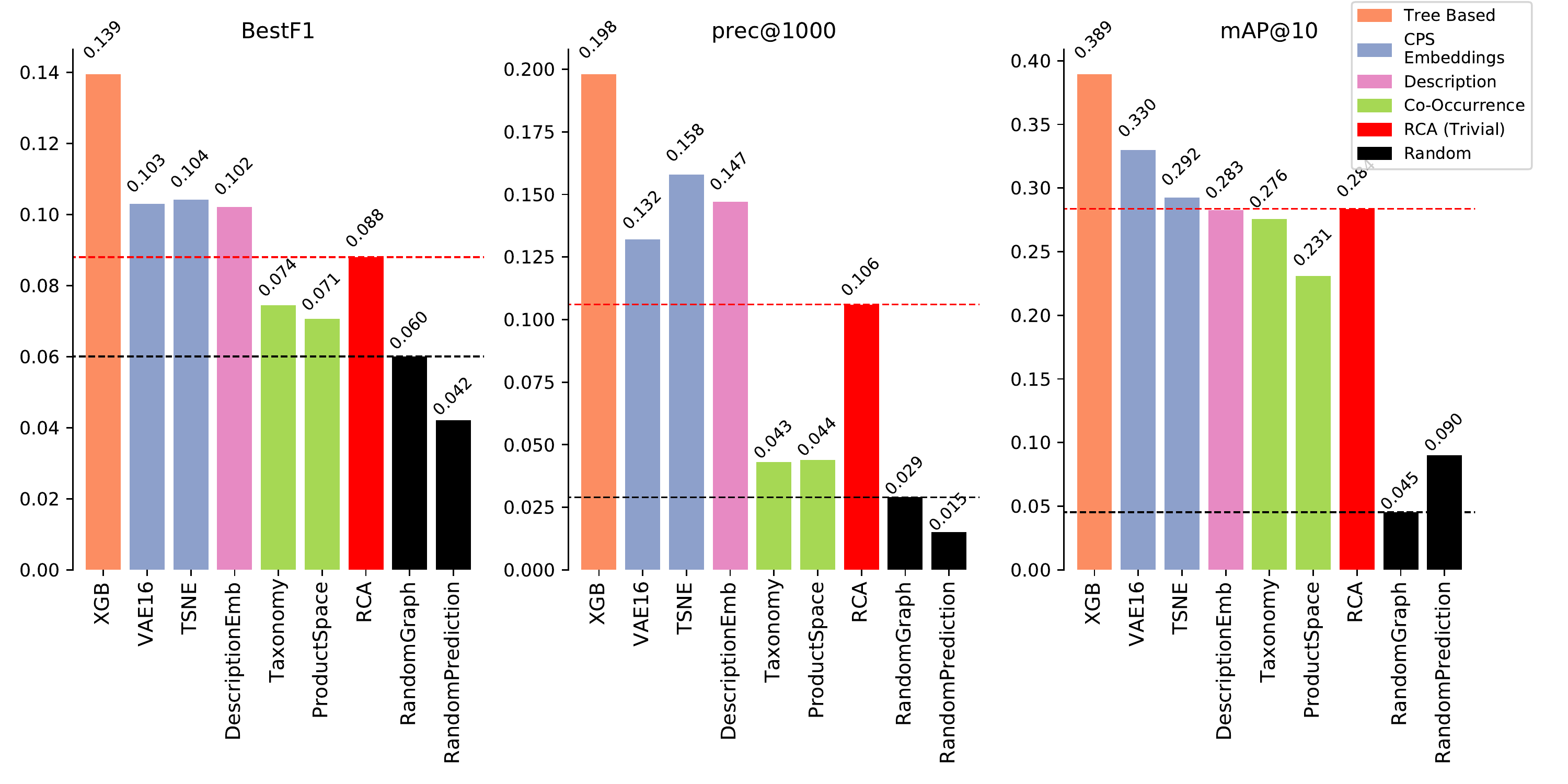}
\caption{Comparison of the prediction performance of different algorithm typologies using different evaluation metrics. Tree-based machine learning techniques (here, XGBoost) overperform all other approaches. Description-based and CPS perform better than the RCA auto-correlation baseline, which in turn overperform co-occurrences.}
\label{fig:distcomp}
\end{figure}

We evaluate the quality of the predictions with several standard classification metrics. In Fig. \ref{fig:distcomp} we show the results for some of these metrics, chosen to be important for practical applications and to capture different aspects of the prediction task. A complete table of the results for all the metrics is available in the SI. The metrics presented in Fig. \ref{fig:distcomp} are (see Methods section for their definitions):

\begin{itemize}
    \item \emph{BestF1}: the F1-Score computed at the optimal decision threshold. This score is computed across all the predicted links for task 2.
    \item \emph{prec@1000}: the precision of the top 1000 predicted links of the $M_{cp}$ matrix in 2018
    \item \emph{mAP@10}: the mean Average Precision of the top 10 predicted links for each country in 2018
\end{itemize}

BestF1 and Precision, therefore, are computed as a function of the predicted links of the whole graph, and we refer to them as \textit{global metrics}, while AP@10 is computed country by country and then averaged (mAP@10). 
\paragraph*{Prediction results}
In the global metrics the co-location based Relatedness proxies perform only marginally better than the Random Graph topology and much worse than the trivial RCA baseline, especially in the prec@1000 metric, where RCA alone performs more than twice as good as Product Space and Taxonomy. In the country-level metric, i.e. the mean Average Precision of top 10 more likely links for each country, the Random Graph performs much worse than the co-location topologies, but the Product Space isn't able to gain any advantage on the trivial RCA prediction, while the Taxonomy manages to score better. The relatively good performance of the Random Graph in the global tasks can be explained with the density-based link prediction approach that we use (see Methods): more diversified countries are more likely to develop new links in the future, and at the same time receive higher scores for all missing links with respect to less diversified countries; this effect is lost in the country-level metrics, that effectively control for diversification by design. It is possible to systematically control for diversification in all density-based prediction strategies (i.e. all the methods presented here with the exception of XGB and RCA) by, e.g. z-scoring all the predictions for each country/year, as suggested in \cite{hidalgo2021economic}. However, as demonstrated by the relatively good results of Random Graph, diversification per-se is a predictive factor and we obtain that about 45\% of the times z-scoring actually worsens the prediction metrics. The full results of these tests are provided in the SI.
Description embeddings and CPS embeddings are able to extract a Relatedness topology that is significantly better in all the global prediction tasks, while in mAP@10 only VAE16 provides a significant improvement over Taxonomy. Both CPS implementations are however always better than the RCA baseline. 
In all tasks, XGB provides a large improvement over all the other pairwise-distances approaches. This is to be expected as the method exploits complex patterns of presence and absence of products in the export baskets of a country and is therefore able to build much more complex decision functions. This comes, of course, at the price of a much less interpretable model. However, the availability of such methods cannot be neglected when the accuracy of the prediction is important, especially for business or policy-relevant decisions.

\begin{figure}[h!]
\centering
\includegraphics[scale=0.74]{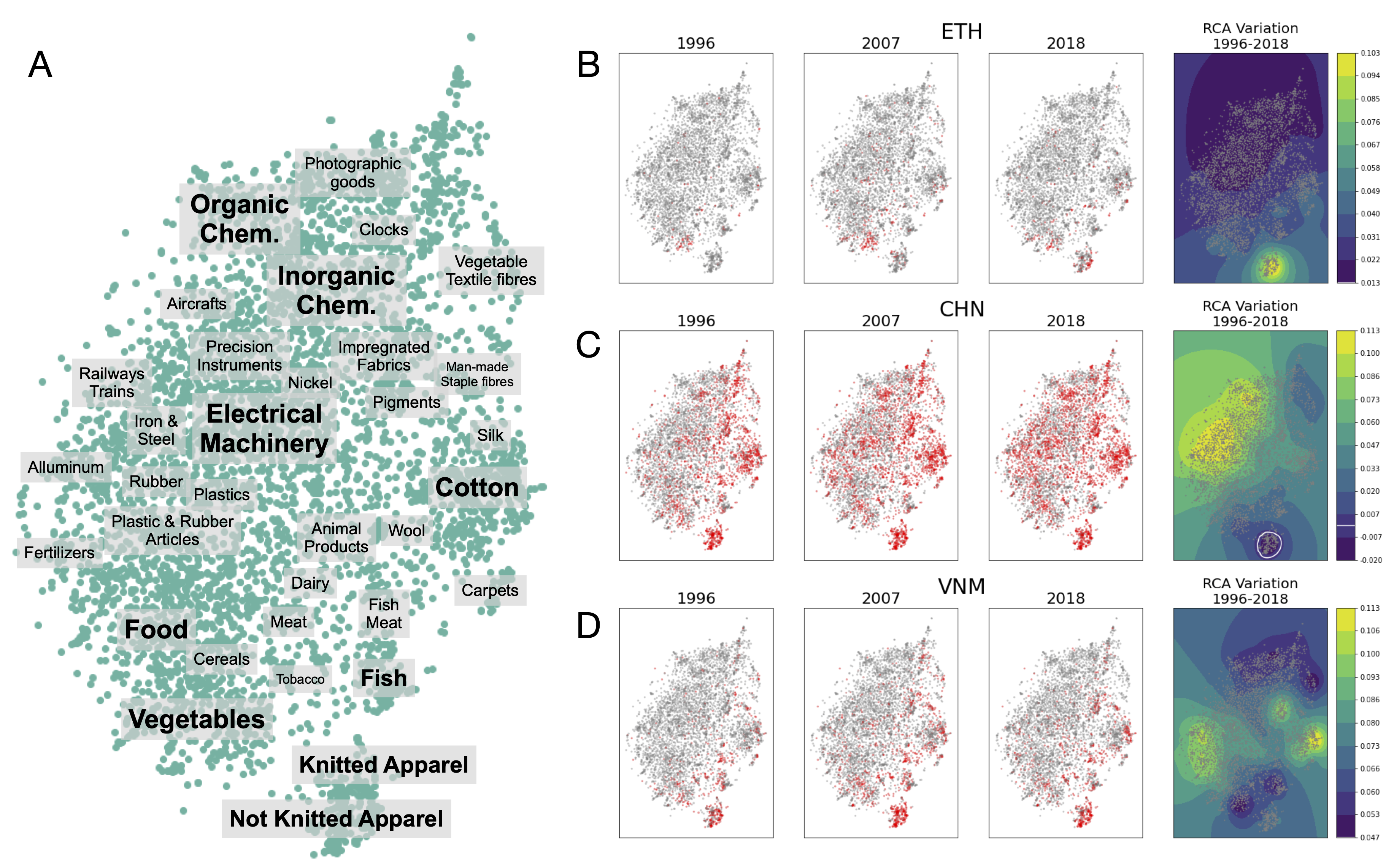}
\caption{\textbf{A.} Two-dimensional CPS/t-SNE embeddings of the Harmonized System 1992 6-digits products. Labels indicate the position of the main clusters. Panels \textbf{B}, \textbf{C} and \textbf{D} show the time evolution of RCA of Ethiopia, China and Vietnam projected on the CPS, with shades of red proportional to the RCA value. The rightmost panels show the average RCA variation averaged with a gaussian kernel. Different development strategies appear very clearly trough the lens of CPS. Ethiopia has strongly focused on the Apparel sector, with a considerably lower increase of RCA in all other sectors. China has focused on heavy industries and had lower RCA growth in other sector, with a negative sign in the Apparel sector. Vietnam shows clearly delimited areas of focus in some heavy industries and textiles, with a much lower growth in agrifood.}
\label{fig:CPS}
\end{figure}
\paragraph*{The Continuous Projection Space}
The advantage of models that predict growth on the basis of pairwise-distances, however, is that they allow to visualize development paths in low dimensional representations of the Relatedness topology. In fig. \ref{fig:CPS} we provide a visualization of the CPS-TSNE embeddings, with the diffusion dynamics of 3 countries on that relatedness topology. For the sake of visualization, the CPS embeddings of fig. \ref{fig:CPS} are computed without cross-validation on data ranging from 1996 to 2018, with products codified in the 1992 version of the Harmonized System. In fig. \ref{fig:CPS} A we label the largest clusters of products that we find, to guide the interpretation of the dynamics highlighted in panels B to C. In those panels we show the diffusion process of 3 countries: Ethiopia, that focused on clothing; China that has strongly diffused towards heavy industries with a net decrease in RCA in clothing; Vietnam has increased its RCA on some heavy industries, although much less than China, and on textiles, with much lower RCA gains in agrifood sectors, that appear to have been deprioritised in Vietnam's strategy. It is interesting to notice how CPS visualisations show at a glance a striking complementarity on the diversification strategies of China and Ethiopia, in a time frame where China has very strongly increased its influence and economic interests in Ethiopia.

\section*{Methods}
\subsection*{Leave-k-countries-out cross-validation}
In this paper we perform link prediction in a temporal bipartite network. The most straightforward way of validating the results out-of-sample would be to learn the forecasting models using the network configurations up to a given time and then evaluate the quality of the forecast on future, unseen data. In the present case however this approach suffers from two shortcomings: (i) the relatively short length of the time-series of data that we consider (12 years) compared to the size of the forecasting window (5 years) implies that we could only learn from a very small portion of the data (7 years), which is insufficient especially for models that explicitly consider the dynamics (e.g. Boosted Trees); (ii) the networks object of this study are extremely auto-correlated in time, as each country tends to change only a very small portion of his product basket from year to year. For this reason, including the past of a country in the training set provides a great amount of information on how that country will look like in the future, that can be directly learned by the model; this is undesired as we want the models to be able to represent the general Relatedness patterns between products, that should not be country-dependent.

To overcome this limitations we adopt a \textit{leave-k-countries-out} approach. We select a set of $k$ countries and we exclude them from the training data. We learn the models on all the available data from the remaining countries. We use the models to predict all the country-product links of the countries left out starting from the year $t_0 + \delta$, where $t_0 = 2007$ in our data and $\delta = 5$. We repeat the process by excluding other subsets of countries until we have a prediction for all countries. Unless otherwise stated, all the cross-validations in this paper have been performed with $k=13$.

\subsection*{Models}
\paragraph*{Notation}
In the following sections we define the models $\mathcal{M}$ as functions that map the export basket of a country $c$ at time $t$, to a score $\mathcal{S}$ proportional to the estimated probability that the country will have $RCA > 1$ in a given product $p$ at time $t+\delta$. We define such models as 
\begin{equation}\label{eq:model}
    \mathcal{M}^{(\Tilde{c})}_{p}(\overrightarrow{M}_c^t) = \mathcal{S}^{t+\delta}_{c,p}
\end{equation}
where $\Tilde{c}$ indicates that the models parameters have been tuned on training data where country $c$ was excluded.

\paragraph*{Density based predictions}
Following the literature\cite{hidalgo2007product}, we define predictive models from the relatedness topologies by considering the density of products in which a country has $RCA > 1$ around the target product $p$ weighted by the relatedness score. More precisely, given a relatedness matrix $B^{(\Tilde{c})}$ computed on data that excludes country $c$, we define
\begin{equation}\label{eq:forcasting_relatedness}
    \mathcal{M}^{(\Tilde{c})}_{p}(\overrightarrow{M}_c^t) = \mathcal{S}^{t+\delta}_{c,p} = \frac{\sum_{p'}B^{(\Tilde{c})}_{p p'} M^t_{c p'}}{\sum_{p'}B^{(\Tilde{c})}_{p p'}}
\end{equation}

\paragraph*{Co-occurence based topologies}
In a seminal paper, Teece et al.\cite{teece1994understanding} introduced the concept of \textit{coherence} between neighboring activities (in their case, products) of firms. In order to measure the relatedness between two activities, they proposed to count the relative co-occurrences, that is the number of firms that are active in both. Using the language of the networks used in Economic Complexity, this corresponds to projecting the bipartite country-product network into a monopartite network of products, computing the product-product adjacency matrix as
\begin{equation} \label{eq:cooccur_teece}
  B_{p p'} = \sum_{c}M_{c p} M_{c p'}. 
\end{equation}
This similarity measure can be normalized in different ways, to take into account trivial effects due to the degree structure of the bipartite network. A more general definition can be written as
\begin{equation}\label{eq:cooccur_gener}
    B_{pp'} = \frac{1}{A}\sum_c\frac{M_{cp}M_{cp'}}{B}
\end{equation}
In this work we consider two specifications: 
\begin{itemize}
    \item \textbf{Product Space:} Setting $A=\max(u_p,u_{p'})$, where $u_p = \sum_c M_{cp}$ is the \textit{ubiquity} of product $p$, and $B=1$ eq.\ref{eq:cooccur_gener} becomes the Product Space \cite{hidalgo2007product}. This normalization controls for the fact that more ubiquitous products have more co-occurrences. 
    \item \textbf{Taxonomy} Setting $A=\max(u_p,u_{p'})$ and $B=d_c$ where $d_c = \sum_p M_{cp}$ is the \textit{diversification} of country $c$, eq.\ref{eq:cooccur_gener} becomes the Taxonomy network \cite{zaccaria2014taxonomy}. This choice of A and B controls again for the ubiquity of products, but also gives a smaller weight to co-occurrences happening in countries with high diversification, as those are more likely to be random.
\end{itemize}
The resulting networks can be filtered using suitable algorithms, such as the Minimal Spanning Tree \cite{hidalgo2007product} or null models \cite{teece1994understanding,cimini2019statistical,pugliese2019unfolding}, but these approaches are beyond the scope of this paper.
When, like in the present case, more than one year of data is available, the relatedness matrix $B$ can be computed by stacking vertically the $M_{cp}$ matrices.


\paragraph*{Product description embeddings}
A relatedness signal can be extracted from the textual description of the products as defined in the Harmonized System Classification. We use this signal as a control for other relatedness measures based on actual data from the countries-products networks. We use the Elmo technique \cite{deep_context_word_repr} to obtain a similarity score between couples of textual descriptions of products. This defines a $B_{pp'}$ relatedness matrix, in analogy with what is done with the co-occurrence based topologies. The forecasting model is built following eq.\ref{eq:forcasting_relatedness}. More details are provided in the SI.

\paragraph*{Boosted Trees}
All the predictive models that we consider in this paper that are based on some form of Relatedness topology, and ultimately make use of eq. \ref{eq:forcasting_relatedness} to build their link prediction score, with all the difference being in how the Relatedness matrix $B_{pp'}$ is built. All these methods, therefore, are based on an independent sum of 2-products relations (Fig. \ref{fig:trees_vs_graphs}A). With decision-tree based methods we can learn more general relations between patterns of presence/absence of sets of products in the export basket of a country, and this can radically improve the link prediction quality (Fig. \ref{fig:trees_vs_graphs}B). 
To learn these complex relationships, however, we are still subject to the same scarcity of data that makes difficult the learning of the co-occurrence based Relatedness matrices. To minimize the risk of learning spurious correlations and to improve the out-of-sample prediction quality, modern decision-tree learning algorithms make use of two ideas: \textit{bagging} and \textit{boosting} (Fig. \ref{fig:trees_vs_graphs}C).
The term \textit{bagging} refers to a data-augmentation strategy where the original training data is randomly sub-sampled many times. In each sub-sample some rows (examples, here countries) and columns (features, here products), randomly chosen, are removed from the training data, and a \textit{weak} model is trained on the remaining data. When the weak models are trees, as in the present case, the column sub-sampling is usually performed randomly every time a new split in the tree is learned. The resulting trained models are defined \textit{weak} as they are learned on incomplete data. However, repeating this operation many times and finally building a meta-model that aggregates the predictions of several weak models is shown to significantly reduce overfitting \cite{friedman2001greedy,freund1997decision}, which would be a significant problem in learning complex models with scarce data.\\
The term \textit{boosting} refers to algorithms that learn weak models in sequence, with each new model trained by giving focus to the training examples that generated the largest losses in the previously learned models. 

The bagging and boosting paradigms can be applied in a variety of ways. Here we make use of the XGBoost \cite{friedman2001greedy,chen2016xgboost} framework to train boosted decision forests. More specifically we train one boosted forest for each product to perform a binary classification task, i.e. we build the input-label pairs as:
\begin{equation}
    (\textit{input}, \textit{label}) = (\overrightarrow{RCA}^t_{c},  M^{t+\delta}_{cp})
\end{equation}
The parameters are the default provided by the XGBoost library (version 1.2.0) except for the \textit{n\_estimators} parameter that has been set to 30. 
To stabilize the results, we repeat the leave-k-countries-out cross-validation 3 times, on 3 different randomizations of the hold-out sets, and average the scores across the 3 runs.
The computational cost of the leave-k-countries-out cross-validation exercise is considerable: in total we train $N_p$ * $N_c/k$ * 3 models, i.e. with $N_p = 5053$, $N_c = 169$ and $k=13$ we train 197067 XGB models.
The boosted trees models perform better than any other model considered in this paper, the main reason being their increased functional complexity. While all the other non-trivial models that we consider are based on binary product-product relationships that are independently summed or averaged together, the boosted trees models explicitly consider higher-order relationships between groups of products. This increased complexity translates into a better capability of the models to represent complex patterns and ultimately to produce better forecasts, but comes at the price of a much lower interpretability.

\begin{figure}[htb]
\centering
\includegraphics[scale=0.74]{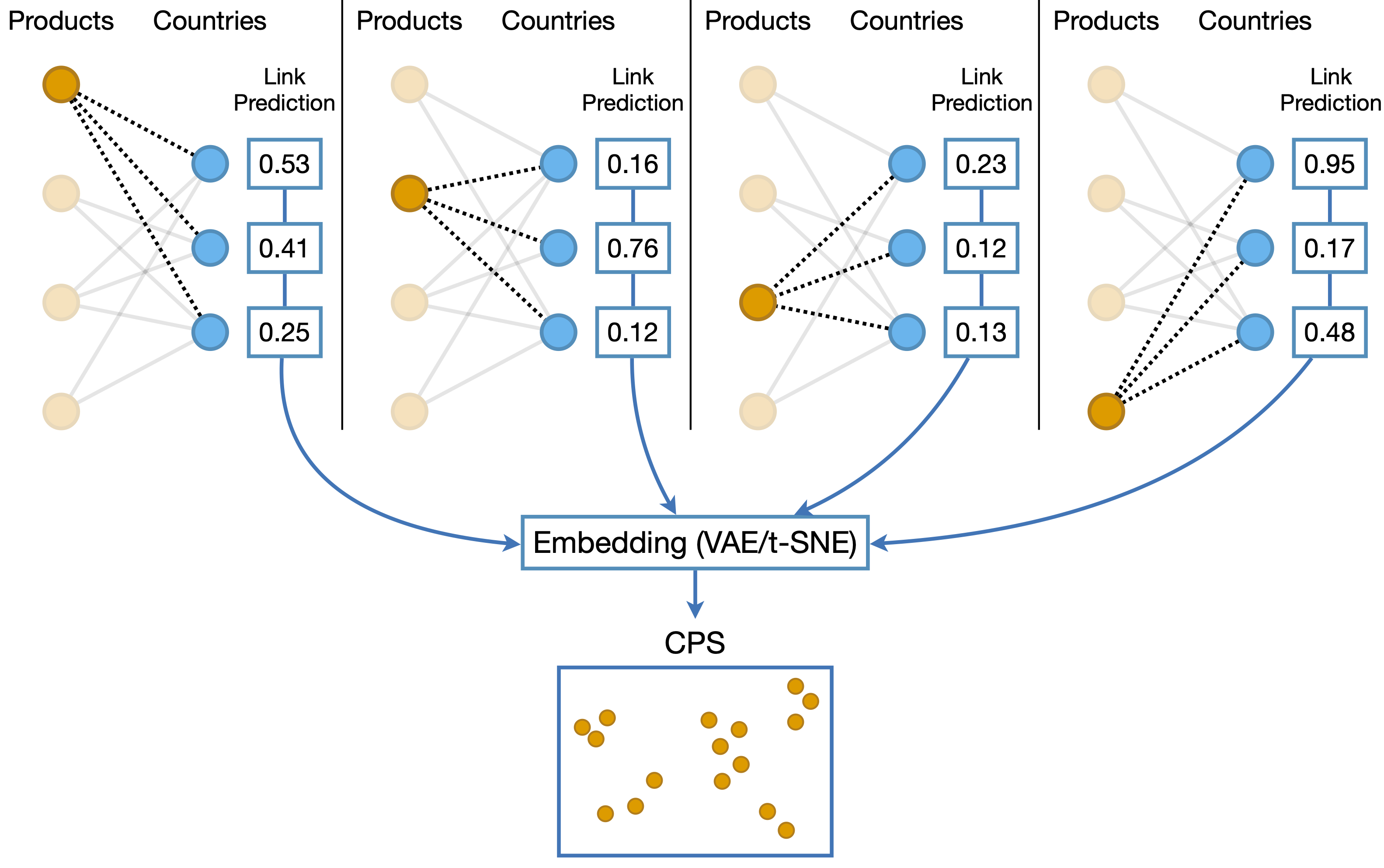}
\caption{Schematic explanation of the CPS methodology. First (top panel) each product is associated with a vector of predictions of the probability to be linked to each country. Then the vectors are embedded with various standard techniques (in this case Variational Autoencoders and t-SNE)}
\label{fig:CPS_methodology}
\end{figure}

\paragraph*{Continuos Projection Space}
We introduce the Continuous Projection Space (CPS) approach in order to recover interpretability from the tree models. The CPS procedure allows to translate the trained tree-based models into a product-product relatedness matrix that retains at the same time the interpretability of co-occurrence based relatedness models and a vastly improved forecasting ability. 
The CPS procedure can be seen as a general self-supervised graph-embedding algorithm, along the lines of several others that gained popularity in the recent literature such as Node2Vec, BiNe and others \cite{goyal2018graph}. The idea behind CPS is to represent each node of a graph through a link-prediction model, i.e. a model that is trained to predict the set of other nodes which that node is linked to and, for weighted graphs, the weight of the link. Once trained, such link prediction model can be \textit{sampled}, by using it to predict links in the graph. In this way we can generate a vector of numbers for each node, i.e. the predicted relation towards each other node in the graph (or a subset of them). This vector embeds information about the model, in terms of his outputs in relation to a fixed set of inputs (i.e. the features used for the link prediction). Intuitively, two nodes that share similar connectivity patterns will be associated to similar models that will, in turn, generate similar vectors of predictions. These vectors can be the CPS embedding per se or, as in the present case, can be further embedded into lower dimensional spaces with general dimensionality reduction techniques. This technique is very general and its effectiveness depends greatly on the choice of the link-prediction models, their training procedure, and the sampling strategy. A general overview and comparison with other approaches in the literature is beyond the scope of this paper, and will be discussed in an upcoming work. Here we provide the specifications of the CPS implementation used in this paper. A schematic explanation of the procedure is provided in fig. \ref{fig:CPS_methodology}

To compute the CPS embeddings in the present bipartite dynamical case, we train Random Forests as the link prediction models to predict 5-years delayed links from each product to the set of countries. That is, we use exactly the same setting that we use to train the boosted trees models, with the same leave-k-countries-out cross validation scheme, the only difference being that we train plain Random Forests instead of Boosted Forests to reduce computational time. After the out-of-sample inference we obtain predictions for all countries, all years and all products, i.e. a tensor $\mathcal{S}$ of the same shape of $M^y_{cp}$, where $y$ indexes the years. Then, for each product, we consider the vector of all the predictions for all the countries and all the years, i.e. a vector of $N_c * N_y$ entries. We then reduce the dimensionality of such vectors in two steps: first we train a 16-dimensional Variational Autoencoder (VAE16) \cite{kingma2013auto}, reducing the vectors to 32 dimensions (i.e. the 16+16 parameters of the VAE), then we perform a further dimensionality reduction from 32 to 2 dimension with the t-SNE algorithm \cite{van2008visualizing}.
To perform the predictions we first compute the $N_p \times N_p$ matrix $D_{pp'}$ of euclidean distances between the embedding vectors. Then we transform $D$ to a matrix of gaussian weights
\begin{equation}\label{eq:gaussian_kernel}
    B_{pp'} = \frac{1}{\sqrt{2\pi\sigma^2}}\exp\left(0.5\left(\frac{D_{pp'}}{\sigma}\right)^2\right)
\end{equation}
Finally we plug such matrix as the $B_{pp'}$ matrix in eq.\ref{eq:forcasting_relatedness} to perform the forecasting. This is equivalent to a Nadaraya-Watson kernel regression with a Gaussian kernel.

It is to be noted that this procedure by itself would not imply a fully out-of-sample prediction that can be directly compared to the results presented for the other methods. This is due to the fact that the leave-k-countries-out cross validation guarantees that the forecasts for each country are done without using any knowledge from that country, but since the CPS embeds each product as a combination of the forecast for all countries then the resulting embeddings actually make use of all the data. For this reason, only for the CPS results, we implement one further step: we compute a set of embeddings and the resulting relatedness matrix $B^{\tilde{c}}_{pp'}$ for each country, by completely eliminating that country from the data, and then using eq.\ref{eq:forcasting_relatedness} to produce forecasts for that country only. The $\sigma$ parameter of eq. \ref{eq:gaussian_kernel} is chosen as the one that maximizes the in-sample Best F1 score in forecasting the links of the countries used to compute $B_{pp'}$.
This procedure is repeated once for each country in the dataset, i.e. 169 times. For this reason, we are not presenting cross-validated CPS results based on Boosted Trees models, but only on the much faster Random Forest.

\paragraph*{RCA Baseline}
As a baseline forecast, we consider the trivial model where
\begin{equation}
    \mathcal{S}^{t+\delta}_{c,p} = \textit{RCA}^t_{c,p}
\end{equation}
Given the strong auto-correlation of the export database, this trivial model provides relatively good predictions. Strikingly, these results overperform the ones obtained from the network of co-occurrences.
\paragraph*{Random Graph Baseline}
The forecasting in eq.\ref{eq:forcasting_relatedness} depends on two terms: one is the Relatedness matrix $B_{pp'}$, the other is the $M_{cp}$ matrix. In the Random Graph baseline model, we randomize the $B_{pp'}$ matrix, i.e. we completely destroy the relatedness signal, but still observe a forecasting power that is better than a completely random forecast and, for some tasks, not far off the co-occurrence based Product Space and Taxonomy matrices. This is due to known \cite{bustos2012dynamics} stylized facts of the country-product matrix, and in particular to its nested structure: more diversified countries are more likely to become even more diversified than non-diversified countries. In eq. \ref{eq:forcasting_relatedness}, even when $B_{pp'}$ is random, more diversified countries get generally higher scores than non diversified ones, even though on random products. This bias is enough to produce forecasts that are better than random and, for some tasks, comparable to co-occurrence based Relatedness topologies.

\subsection*{Evaluation metrics}
In order to compare the different prediction methodologies we make use of a series of performance indicators, usually adopted for classification tasks. It is important to point out that our results (see for instance fig.\ref{fig:trees_vs_graphs}) are highly consistent across different indicators, even if we choose them for covering different aspects of the prediction exercise. Let us now discuss in detail how the prediction performances can be quantitatively evaluated. The specific instance to predict can be \textit{positive} or \textit{negative}, if the corresponding element of the export matrix $M_{cp}$ is equal to 1 or 0, respectively. A \textit{true positive} is a correctly predicted positive instance. Let us focus on the top $k=1000$ scores of a given algorithm, that is, this algorithm assigns to these couples $(c,p)$ the higher likelihoods to have $M_{cp}=1$. The indicator \textit{prec@1000} is defined as the fraction of these 1000 couples for which $M_{cp}=1$. This is a measure of the global precision of the algorithm. However, this measure takes into account all matrix elements together, while we might be interested in evaluating the prediction performance on a country basis: on average, how much are we precise when recommending a product to a country? To do so we first evaluate the precision country by country, and then we average. This is called the mean precision. Moreover, it is also important to weight our success or failure using the scores rank: we want the highest scores to predict better than the lowest scores. So we have to compute a weighted average. In practice, we use the mean Average Precision \textit{mAP@10}. Let us focus on a single country $c$ first. The Average Precision \textit{AP@k(c)} is defined as
\begin{equation}
    AP@k(c)=\frac{\sum_{i=1}^{k}prec@k(c)\times rel(k,c)}{\min(k,P(c))}
\end{equation}
where \textit(prec@k(c)) is the precision at k for country c in the list; rel(k,c) is equal to 1 if the item at rank k is positive and zero otherwise; and P(c) is the total number of positives for country c. Then, the \textit{mAP@k} is simply the country average:  
\begin{equation}
    mAP@k=\frac{\sum_{c=1}^{C}AP@k(c)}{C}
\end{equation}
Precision-related measures deal with the minimization of false positives FP. In order to take into account also the problem of false negatives FN, \textit{recall} is usually considered. In general, precision is defined as the ratio between the number of true positives $TP$ and the number of predicted positives, while recall is the ratio between true positives $TP$ and true instances, the real positives $P$. Since we want a global and balanced measure, we average the two with an harmonic mean, called F1-score. The harmonic mean aggravates the impact of possible small values of one of the two indicators. Since binary classifiers usually provide a continuous set of scores, one has to specify a threshold $t$ above which the score is associated to a positive prediction; as a consequence, precision, recall, and the F1 score will depend on $t$. We point out that in the computation of \textit{prec@1000} the threshold choice is derived from the choice $k=1000$. To have a non trivial and non arbitrary threshold we decided to take the one that maximizes the in-sample F1 score, as suggested by \cite{manning2008introduction}. Summarizing the above considerations, the best F1 score shown in Fig.\ref{fig:trees_vs_graphs} is defined as
\begin{equation}
    \text{BestF1score}=\max_{t} \frac{2}{\text{prec(t)}^{-1}+\text{rec(t)}^{-1}}
\end{equation}
where
\begin{equation}
    \text{prec(t)}=\frac{TP(t)}{TP(t)+FP(t)} \qquad \text{rec(t)}=\frac{TP(t)}{TP(t)+FN(t)}=\frac{TP(t)}{P}.
\end{equation}
These prediction performance measures can be computed for any test set. In order to show the replicability and the extent of our results, we show in the SI that they do not change if the test set and the indicators are reasonably changed. In Fig.\ref{fig:trees_vs_graphs} we compute the \textit{prec@1000} and the \textit{mAP@10} for the last year of our dataset (2018): this bears the interpretation of our result as a recommender system, in which products are suggested as feasible to countries, and countries actually start exporting them. Instead, the best F1-score is computed on all the available years in cross validation, to show that our results are stable and comparable across different periods.

\subsection*{Trade data}\label{sec:trade_data}
The UN-COMTRADE database (https://comtrade.un.org) provides the monetary volumes of the trade flows between countries. Strikingly, the information is provided on a product basis, so we know which country is exporting a given amount of a given product. Since importers' and exporters' declarations do not coincide, suitable reconstruction algorithms are needed in order to achieve a coherent and sanitized dataset. By using a global Bayesian optimization approach, we produced a denoised dataset \cite{mazzilli2021reconstruct} that permits, by the way, to increase the GDP prediction performance in a considerable way \cite{tacchella2018dynamical}. 
\\

\section*{Discussion}
The concept of Relatedness can be an extremely powerful tool to understand development dynamics and to inform policy decisions. By quantifying the proximity between activities in terms of knowledge, inputs, and infrastructures, it can help designing paths to diversification or specialization strategies, based on empirical evidence. However, while the idea is extremely appealing, we show that current methods to estimate Relatedness from data perform poorly in predicting the trajectories of countries, even when compared to trivial alternative approaches. In order to overcome these limitations we have introduced a novel network embedding technique called Continuous Projection Space (CPS) for the computation of Relatedness from data. CPS performs up to 2.6 times better than current approaches while retaining the same overall properties and interpretability. Moreover, we have shown that moving from one-to-one product Relatedness to many-to-one relationships using suitable Machine Learning algorithms allows to achieve performances up to 3.5 times better than current approaches, despite loosing some of their interpretability.\\
We believe that these results can have a huge impact on the applicability of these ideas in policy-making as the quality and confidence of the recommendations are dramatically improved. The magnitude of the improvement, especially in many-to-one models, is enough to concretely move these methods from a research idea to an applied tool to inform policy decisions. The adoption has already started in large institutions such as the World Bank and the European Commission.\\
CPS, despite having a relatively lower performance with respect to the many-to-one models, has the advantage of providing a fully explainable prediction in terms of pairwise euclidean distances, which can be of practical interest for policy makers. Besides allowing to visualize and explore the Relatedness space, this fact can have the added benefit of generalization. It would be technically possible to use the same Relatedness metrics learned at the national level to inform development strategies in regions, cities, firms, all entities for which obtaining a consistent worldwide dataset would be extremely harder. This is due to the linear, additive form of eq.\ref{eq:forcasting_relatedness}. In contrast, this generalization is much less likely to work in tree based methods.\\
The work presented here opens to various further research ideas, of which at least two deserve to be mentioned. First, the CPS approach is a general Network Embedding technique that can in principle be applied to monopartite networks as well as bipartite or multipartite networks, as in the present case. We plan to systematically explore the capabilities of the CPS technique and compare it with the existing literature\footnote{In the present case a comparison with the state of the art method for embedding bipartite networks, BiNE, is provided in the SI. Its performance is much lower than CPS.}. Second, the tree-based prediction is suited to be generalized to multi-partite networks as well (such as, e.g., the Countries-Technologies-Products-Research network used in \cite{pugliese2019unfolding}), and we plan to explore if mixing information from multiple layers can indeed improve the quality of the predictions. Finally, a CPS embedding is in theory feasible also for such multipartite networks, allowing to embed nodes from different layers in a common space.

\bibliography{scibib}

\begin{thebibliography}{10}
\expandafter\ifx\csname url\endcsname\relax
  \def\url#1{\texttt{#1}}\fi
\expandafter\ifx\csname urlprefix\endcsname\relax\def\urlprefix{URL }\fi
\providecommand{\bibinfo}[2]{#2}
\providecommand{\eprint}[2][]{\url{#2}}

\bibitem{hidalgo2018principle}
\bibinfo{author}{Hidalgo, C.~A.} \emph{et~al.}
\newblock \bibinfo{title}{The principle of relatedness}.
\newblock In \emph{\bibinfo{booktitle}{International conference on complex
  systems}}, \bibinfo{pages}{451--457} (\bibinfo{organization}{Springer},
  \bibinfo{year}{2018}).

\bibitem{balland2019smart}
\bibinfo{author}{Balland, P.-A.}, \bibinfo{author}{Boschma, R.},
  \bibinfo{author}{Crespo, J.} \& \bibinfo{author}{Rigby, D.~L.}
\newblock \bibinfo{title}{Smart specialization policy in the european union:
  relatedness, knowledge complexity and regional diversification}.
\newblock \emph{\bibinfo{journal}{Regional Studies}}
  \textbf{\bibinfo{volume}{53}}, \bibinfo{pages}{1252--1268}
  (\bibinfo{year}{2019}).

\bibitem{hidalgo2007product}
\bibinfo{author}{Hidalgo, C.~A.}, \bibinfo{author}{Klinger, B.},
  \bibinfo{author}{Barab{\'a}si, A.-L.} \& \bibinfo{author}{Hausmann, R.}
\newblock \bibinfo{title}{The product space conditions the development of
  nations}.
\newblock \emph{\bibinfo{journal}{Science}} \textbf{\bibinfo{volume}{317}},
  \bibinfo{pages}{482--487} (\bibinfo{year}{2007}).

\bibitem{zaccaria2014taxonomy}
\bibinfo{author}{Zaccaria, A.}, \bibinfo{author}{Cristelli, M.},
  \bibinfo{author}{Tacchella, A.} \& \bibinfo{author}{Pietronero, L.}
\newblock \bibinfo{title}{How the taxonomy of products drives the economic
  development of countries}.
\newblock \emph{\bibinfo{journal}{PloS one}} \textbf{\bibinfo{volume}{9}},
  \bibinfo{pages}{e113770} (\bibinfo{year}{2014}).

\bibitem{lu2012recommender}
\bibinfo{author}{L{\"u}, L.} \emph{et~al.}
\newblock \bibinfo{title}{Recommender systems}.
\newblock \emph{\bibinfo{journal}{Physics reports}}
  \textbf{\bibinfo{volume}{519}}, \bibinfo{pages}{1--49}
  (\bibinfo{year}{2012}).

\bibitem{teece1994understanding}
\bibinfo{author}{Teece, D.~J.}, \bibinfo{author}{Rumelt, R.},
  \bibinfo{author}{Dosi, G.} \& \bibinfo{author}{Winter, S.}
\newblock \bibinfo{title}{Understanding corporate coherence: Theory and
  evidence}.
\newblock \emph{\bibinfo{journal}{Journal of economic behavior \&
  organization}} \textbf{\bibinfo{volume}{23}}, \bibinfo{pages}{1--30}
  (\bibinfo{year}{1994}).

\bibitem{hidalgo2021economic}
\bibinfo{author}{Hidalgo, C.~A.}
\newblock \bibinfo{title}{Economic complexity theory and applications}.
\newblock \emph{\bibinfo{journal}{Nature Reviews Physics}}
  \bibinfo{pages}{1--22} (\bibinfo{year}{2021}).

\bibitem{hidalgo2009building}
\bibinfo{author}{Hidalgo, C.~A.} \& \bibinfo{author}{Hausmann, R.}
\newblock \bibinfo{title}{The building blocks of economic complexity}.
\newblock \emph{\bibinfo{journal}{proceedings of the national academy of
  sciences}} \textbf{\bibinfo{volume}{106}}, \bibinfo{pages}{10570--10575}
  (\bibinfo{year}{2009}).

\bibitem{Tacchella}
\bibinfo{author}{Tacchella, A.}, \bibinfo{author}{Cristelli, M.},
  \bibinfo{author}{Caldarelli, G.}, \bibinfo{author}{Gabrielli, A.} \&
  \bibinfo{author}{Pietronero, L.}
\newblock \bibinfo{title}{A new metrics for countries' fitness and products'
  complexity}.
\newblock \emph{\bibinfo{journal}{Scientific Reports}}
  \textbf{\bibinfo{volume}{2}}, \bibinfo{pages}{723} (\bibinfo{year}{2012}).

\bibitem{tacchella2013economic}
\bibinfo{author}{Tacchella, A.}, \bibinfo{author}{Cristelli, M.},
  \bibinfo{author}{Caldarelli, G.}, \bibinfo{author}{Gabrielli, A.} \&
  \bibinfo{author}{Pietronero, L.}
\newblock \bibinfo{title}{Economic complexity: conceptual grounding of a new
  metrics for global competitiveness}.
\newblock \emph{\bibinfo{journal}{Journal of Economic Dynamics and Control}}
  \textbf{\bibinfo{volume}{37}}, \bibinfo{pages}{1683--1691}
  (\bibinfo{year}{2013}).

\bibitem{cristelli2013measuring}
\bibinfo{author}{Cristelli, M.}, \bibinfo{author}{Gabrielli, A.},
  \bibinfo{author}{Tacchella, A.}, \bibinfo{author}{Caldarelli, G.} \&
  \bibinfo{author}{Pietronero, L.}
\newblock \bibinfo{title}{Measuring the intangibles: A metrics for the economic
  complexity of countries and products}.
\newblock \emph{\bibinfo{journal}{PloS one}} \textbf{\bibinfo{volume}{8}},
  \bibinfo{pages}{e70726} (\bibinfo{year}{2013}).

\bibitem{sbardella2018role}
\bibinfo{author}{Sbardella, A.}, \bibinfo{author}{Pugliese, E.},
  \bibinfo{author}{Zaccaria, A.} \& \bibinfo{author}{Scaramozzino, P.}
\newblock \bibinfo{title}{The role of complex analysis in modelling economic
  growth}.
\newblock \emph{\bibinfo{journal}{Entropy}} \textbf{\bibinfo{volume}{20}},
  \bibinfo{pages}{883} (\bibinfo{year}{2018}).

\bibitem{lin2020african}
\bibinfo{author}{Lin, J.}, \bibinfo{author}{Cader, M.} \&
  \bibinfo{author}{Pietronero, L.}
\newblock \bibinfo{title}{What african industrial development can learn from
  east asian successes}  (\bibinfo{year}{2020}).

\bibitem{pugliese2019economic}
\bibinfo{author}{Pugliese, E.} \& \bibinfo{author}{Tübke, A.}
\newblock \bibinfo{title}{Economic complexity to address current challenges in
  innovation systems: A novel empirical strategy linked to the territorial
  dimension}.
\newblock \emph{\bibinfo{journal}{Industrial R\&I –JRC Policy Insights}}
  (\bibinfo{year}{2019}).

\bibitem{pugliese2019economic2}
\bibinfo{author}{Pugliese, E.} \& \bibinfo{author}{Tacchella, A.}
\newblock \bibinfo{title}{Economic complexity for competitiveness and
  innovation: a novel bottom-up strategy linking global and regional
  capacities}.
\newblock \emph{\bibinfo{journal}{Industrial R\&I –JRC Policy Insights}}
  (\bibinfo{year}{2020}).

\bibitem{van2020vertical}
\bibinfo{author}{van Dam, A.} \& \bibinfo{author}{Frenken, K.}
\newblock \bibinfo{title}{Vertical vs. horizontal policy in a capabilities
  model of economic development}.
\newblock \emph{\bibinfo{journal}{arXiv preprint arXiv:2006.04624}}
  (\bibinfo{year}{2020}).

\bibitem{tacchella2016build}
\bibinfo{author}{Tacchella, A.}, \bibinfo{author}{Di~Clemente, R.},
  \bibinfo{author}{Gabrielli, A.} \& \bibinfo{author}{Pietronero, L.}
\newblock \bibinfo{title}{The build-up of diversity in complex ecosystems}.
\newblock \emph{\bibinfo{journal}{arXiv preprint arXiv:1609.03617}}
  (\bibinfo{year}{2016}).

\bibitem{neffke2011regions}
\bibinfo{author}{Neffke, F.}, \bibinfo{author}{Henning, M.} \&
  \bibinfo{author}{Boschma, R.}
\newblock \bibinfo{title}{How do regions diversify over time? industry
  relatedness and the development of new growth paths in regions}.
\newblock \emph{\bibinfo{journal}{Economic geography}}
  \textbf{\bibinfo{volume}{87}}, \bibinfo{pages}{237--265}
  (\bibinfo{year}{2011}).

\bibitem{saracco2017inferring}
\bibinfo{author}{Saracco, F.} \emph{et~al.}
\newblock \bibinfo{title}{Inferring monopartite projections of bipartite
  networks: an entropy-based approach}.
\newblock \emph{\bibinfo{journal}{New Journal of Physics}}
  \textbf{\bibinfo{volume}{19}}, \bibinfo{pages}{053022}
  (\bibinfo{year}{2017}).

\bibitem{pugliese2019unfolding}
\bibinfo{author}{Pugliese, E.} \emph{et~al.}
\newblock \bibinfo{title}{Unfolding the innovation system for the development
  of countries: coevolution of science, technology and production}.
\newblock \emph{\bibinfo{journal}{Scientific reports}}
  \textbf{\bibinfo{volume}{9}}, \bibinfo{pages}{1--12} (\bibinfo{year}{2019}).

\bibitem{zaccaria2018integrating}
\bibinfo{author}{Zaccaria, A.}, \bibinfo{author}{Mishra, S.},
  \bibinfo{author}{Cader, M.~Z.} \& \bibinfo{author}{Pietronero, L.}
\newblock \bibinfo{title}{Integrating services in the economic fitness
  approach}.
\newblock \emph{\bibinfo{journal}{World Bank Policy Research Working Paper}}
  (\bibinfo{year}{2018}).

\bibitem{stojkoski2016impact}
\bibinfo{author}{Stojkoski, V.}, \bibinfo{author}{Utkovski, Z.} \&
  \bibinfo{author}{Kocarev, L.}
\newblock \bibinfo{title}{The impact of services on economic complexity:
  Service sophistication as route for economic growth}.
\newblock \emph{\bibinfo{journal}{PloS one}} \textbf{\bibinfo{volume}{11}}
  (\bibinfo{year}{2016}).

\bibitem{bun2017cleaning}
\bibinfo{author}{Bun, J.}, \bibinfo{author}{Bouchaud, J.-P.} \&
  \bibinfo{author}{Potters, M.}
\newblock \bibinfo{title}{Cleaning large correlation matrices: tools from
  random matrix theory}.
\newblock \emph{\bibinfo{journal}{Physics Reports}}
  \textbf{\bibinfo{volume}{666}}, \bibinfo{pages}{1--109}
  (\bibinfo{year}{2017}).

\bibitem{mariani2019nestedness}
\bibinfo{author}{Mariani, M.~S.}, \bibinfo{author}{Ren, Z.-M.},
  \bibinfo{author}{Bascompte, J.} \& \bibinfo{author}{Tessone, C.~J.}
\newblock \bibinfo{title}{Nestedness in complex networks: observation,
  emergence, and implications}.
\newblock \emph{\bibinfo{journal}{Physics Reports}}
  \textbf{\bibinfo{volume}{813}}, \bibinfo{pages}{1--90}
  (\bibinfo{year}{2019}).

\bibitem{nesta2005coherence}
\bibinfo{author}{Nesta, L.} \& \bibinfo{author}{Saviotti, P.~P.}
\newblock \bibinfo{title}{Coherence of the knowledge base and the firm's
  innovative performance: evidence from the us pharmaceutical industry}.
\newblock \emph{\bibinfo{journal}{The Journal of Industrial Economics}}
  \textbf{\bibinfo{volume}{53}}, \bibinfo{pages}{123--142}
  (\bibinfo{year}{2005}).

\bibitem{bottazzi2010measuring}
\bibinfo{author}{Bottazzi, G.} \& \bibinfo{author}{Pirino, D.}
\newblock \bibinfo{title}{Measuring industry relatedness and corporate
  coherence}.
\newblock \emph{\bibinfo{journal}{Available at SSRN 1831479}}
  (\bibinfo{year}{2010}).

\bibitem{li2014statistically}
\bibinfo{author}{Li, M.-X.} \emph{et~al.}
\newblock \bibinfo{title}{Statistically validated mobile communication
  networks: the evolution of motifs in european and chinese data}.
\newblock \emph{\bibinfo{journal}{New Journal of Physics}}
  \textbf{\bibinfo{volume}{16}}, \bibinfo{pages}{083038}
  (\bibinfo{year}{2014}).

\bibitem{cimini2019statistical}
\bibinfo{author}{Cimini, G.} \emph{et~al.}
\newblock \bibinfo{title}{The statistical physics of real-world networks}.
\newblock \emph{\bibinfo{journal}{Nature Reviews Physics}}
  \textbf{\bibinfo{volume}{1}}, \bibinfo{pages}{58--71} (\bibinfo{year}{2019}).

\bibitem{giorgio2020knowledge}
\bibinfo{author}{Tripodi, G.}, \bibinfo{author}{Chiaromonte, F.} \&
  \bibinfo{author}{Lillo, F.}
\newblock \bibinfo{title}{Knowledge and social relatedness shape research
  portfolio diversification}.
\newblock \emph{\bibinfo{journal}{Scientific Reports (Nature Publisher Group)}}
  \textbf{\bibinfo{volume}{10}} (\bibinfo{year}{2020}).

\bibitem{zhou2010solving}
\bibinfo{author}{Zhou, T.} \emph{et~al.}
\newblock \bibinfo{title}{Solving the apparent diversity-accuracy dilemma of
  recommender systems}.
\newblock \emph{\bibinfo{journal}{Proceedings of the National Academy of
  Sciences}} \textbf{\bibinfo{volume}{107}}, \bibinfo{pages}{4511--4515}
  (\bibinfo{year}{2010}).

\bibitem{zhou2007bipartite}
\bibinfo{author}{Zhou, T.}, \bibinfo{author}{Ren, J.}, \bibinfo{author}{Medo,
  M.} \& \bibinfo{author}{Zhang, Y.-C.}
\newblock \bibinfo{title}{Bipartite network projection and personal
  recommendation}.
\newblock \emph{\bibinfo{journal}{Physical review E}}
  \textbf{\bibinfo{volume}{76}}, \bibinfo{pages}{046115}
  (\bibinfo{year}{2007}).

\bibitem{che2020intelligent}
\bibinfo{author}{Che, N.~X.}
\newblock \bibinfo{title}{Intelligent export diversification: An export
  recommendation system with machine learning}.
\newblock \bibinfo{type}{Tech. Rep.}, \bibinfo{institution}{International
  Monetary Fund} (\bibinfo{year}{2020}).

\bibitem{tacchella2020language}
\bibinfo{author}{Tacchella, A.}, \bibinfo{author}{Napoletano, A.} \&
  \bibinfo{author}{Pietronero, L.}
\newblock \bibinfo{title}{The language of innovation}.
\newblock \emph{\bibinfo{journal}{PloS one}} \textbf{\bibinfo{volume}{15}},
  \bibinfo{pages}{e0230107} (\bibinfo{year}{2020}).

\bibitem{palmucci2020your}
\bibinfo{author}{Palmucci, A.}, \bibinfo{author}{Liao, H.},
  \bibinfo{author}{Napoletano, A.} \& \bibinfo{author}{Zaccaria, A.}
\newblock \bibinfo{title}{Where is your field going? a machine learning
  approach to study the relative motion of the domains of physics}.
\newblock \emph{\bibinfo{journal}{PloS one}} \textbf{\bibinfo{volume}{15}},
  \bibinfo{pages}{e0233997} (\bibinfo{year}{2020}).

\bibitem{balassa1965trade}
\bibinfo{author}{Balassa, B.}
\newblock \bibinfo{title}{Trade liberalisation and “revealed” comparative
  advantage 1}.
\newblock \emph{\bibinfo{journal}{The manchester school}}
  \textbf{\bibinfo{volume}{33}}, \bibinfo{pages}{99--123}
  (\bibinfo{year}{1965}).

\bibitem{alshamsi2018optimal}
\bibinfo{author}{Alshamsi, A.}, \bibinfo{author}{Pinheiro, F.~L.} \&
  \bibinfo{author}{Hidalgo, C.~A.}
\newblock \bibinfo{title}{Optimal diversification strategies in the networks of
  related products and of related research areas}.
\newblock \emph{\bibinfo{journal}{Nature communications}}
  \textbf{\bibinfo{volume}{9}}, \bibinfo{pages}{1--7} (\bibinfo{year}{2018}).

\bibitem{fuzzyset}
\bibinfo{author}{Zhelezniak, V.} \emph{et~al.}
\newblock \bibinfo{title}{Don't settle for average, go for the max: Fuzzy sets
  and max-pooled word vectors}.
\newblock \emph{\bibinfo{journal}{arXiv preprint arXiv:1904.13264}}
  (\bibinfo{year}{2019}).

\bibitem{friedman2001greedy}
\bibinfo{author}{Friedman, J.~H.}
\newblock \bibinfo{title}{Greedy function approximation: a gradient boosting
  machine}.
\newblock \emph{\bibinfo{journal}{Annals of statistics}}
  \bibinfo{pages}{1189--1232} (\bibinfo{year}{2001}).

\bibitem{chen2016xgboost}
\bibinfo{author}{Chen, T.} \& \bibinfo{author}{Guestrin, C.}
\newblock \bibinfo{title}{Xgboost: A scalable tree boosting system}.
\newblock In \emph{\bibinfo{booktitle}{Proceedings of the 22nd acm sigkdd
  international conference on knowledge discovery and data mining}},
  \bibinfo{pages}{785--794} (\bibinfo{year}{2016}).

\bibitem{van2008visualizing}
\bibinfo{author}{Van~der Maaten, L.} \& \bibinfo{author}{Hinton, G.}
\newblock \bibinfo{title}{Visualizing data using t-sne.}
\newblock \emph{\bibinfo{journal}{Journal of machine learning research}}
  \textbf{\bibinfo{volume}{9}} (\bibinfo{year}{2008}).

\bibitem{kingma2013auto}
\bibinfo{author}{Kingma, D.~P.} \& \bibinfo{author}{Welling, M.}
\newblock \bibinfo{title}{Auto-encoding variational bayes}.
\newblock \emph{\bibinfo{journal}{arXiv preprint arXiv:1312.6114}}
  (\bibinfo{year}{2013}).

\bibitem{deep_context_word_repr}
\bibinfo{author}{Peters, M.~E.} \emph{et~al.}
\newblock \bibinfo{title}{Deep contextualized word representations}.
\newblock \emph{\bibinfo{journal}{arXiv preprint arXiv:1802.05365}}
  (\bibinfo{year}{2018}).

\bibitem{freund1997decision}
\bibinfo{author}{Freund, Y.} \& \bibinfo{author}{Schapire, R.~E.}
\newblock \bibinfo{title}{A decision-theoretic generalization of on-line
  learning and an application to boosting}.
\newblock \emph{\bibinfo{journal}{Journal of computer and system sciences}}
  \textbf{\bibinfo{volume}{55}}, \bibinfo{pages}{119--139}
  (\bibinfo{year}{1997}).

\bibitem{goyal2018graph}
\bibinfo{author}{Goyal, P.} \& \bibinfo{author}{Ferrara, E.}
\newblock \bibinfo{title}{Graph embedding techniques, applications, and
  performance: A survey}.
\newblock \emph{\bibinfo{journal}{Knowledge-Based Systems}}
  \textbf{\bibinfo{volume}{151}}, \bibinfo{pages}{78--94}
  (\bibinfo{year}{2018}).

\bibitem{bustos2012dynamics}
\bibinfo{author}{Bustos, S.}, \bibinfo{author}{Gomez, C.},
  \bibinfo{author}{Hausmann, R.} \& \bibinfo{author}{Hidalgo, C.~A.}
\newblock \bibinfo{title}{The dynamics of nestedness predicts the evolution of
  industrial ecosystems}.
\newblock \emph{\bibinfo{journal}{PloS one}} \textbf{\bibinfo{volume}{7}},
  \bibinfo{pages}{e49393} (\bibinfo{year}{2012}).

\bibitem{manning2008introduction}
\bibinfo{author}{Manning, C.~D.}, \bibinfo{author}{Raghavan, P.} \&
  \bibinfo{author}{Sch{\"u}tze, H.}
\newblock \emph{\bibinfo{title}{Introduction to information retrieval}}
  (\bibinfo{publisher}{Cambridge university press}, \bibinfo{year}{2008}).

\bibitem{mazzilli2021reconstruct}
\bibinfo{author}{Mazzilli, D.}, \bibinfo{author}{Andrea, T.} \&
  \bibinfo{author}{Pietronero, L.}
\newblock \bibinfo{title}{In preparation}  (\bibinfo{year}{2021}).

\bibitem{tacchella2018dynamical}
\bibinfo{author}{Tacchella, A.}, \bibinfo{author}{Mazzilli, D.} \&
  \bibinfo{author}{Pietronero, L.}
\newblock \bibinfo{title}{A dynamical systems approach to gross domestic
  product forecasting}.
\newblock \emph{\bibinfo{journal}{Nature Physics}}
  \textbf{\bibinfo{volume}{14}}, \bibinfo{pages}{861--865}
  (\bibinfo{year}{2018}).

\end{thebibliography}

\bibliographystyle{naturemag}


\clearpage

\end{document}